\documentclass[manuscript]{aastex63}

\usepackage[utf8]{inputenc}
\usepackage{natbib}
\usepackage{color}
\usepackage{graphicx}
\usepackage{rotating}
\usepackage{hyperref}
\usepackage[all]{hypcap}
\usepackage{mathrsfs}
\usepackage{amsmath}
\usepackage{amssymb}
\usepackage{float}
\usepackage{graphicx}
\usepackage{subfigure}
\usepackage{lineno}

\hypersetup{
    unicode=false,          
    pdftoolbar=true,        
    pdfmenubar=true,        
    pdffitwindow=false,     
    pdfstartview={FitH},    
    pdftitle={My title},    
    pdfauthor={Author},     
    pdfsubject={Subject},   
    pdfcreator={Creator},   
    pdfproducer={Producer}, 
    pdfkeywords={keyword1} {key2} {key3}, 
    pdfnewwindow=true,      
    colorlinks=true,        
    linkcolor=blue,         
    citecolor=blue,         
    filecolor=magenta,      
    urlcolor=cyan           
}

\newcommand{\p}[1]{{\color{red}{#1}}}

\begin{document}
\title{Data-constrained Magnetohydrodynamic Simulation of an Intermediate Solar Filament Eruption}

\author[0000-0002-9293-8439]{Yang Guo}
\affil{School of Astronomy and Space Science and Key Laboratory for Modern Astronomy and Astrophysics, Nanjing University, Nanjing 210023, China} \email{guoyang@nju.edu.cn}

\author{Jinhan Guo}
\affil{School of Astronomy and Space Science and Key Laboratory for Modern Astronomy and Astrophysics, Nanjing University, Nanjing 210023, China} 
\affil{Centre for mathematical Plasma Astrophysics, Department of Mathematics, KU Leuven, Celestijnenlaan 200B, B-3001 Leuven, Belgium}

\author{Yiwei Ni}
\affil{School of Astronomy and Space Science and Key Laboratory for Modern Astronomy and Astrophysics, Nanjing University, Nanjing 210023, China} 

\author[0000-0002-4978-4972]{M. D. Ding}
\affil{School of Astronomy and Space Science and Key Laboratory for Modern Astronomy and Astrophysics, Nanjing University, Nanjing 210023, China} 

\author[0000-0002-7289-642X]{P. F. Chen}
\affil{School of Astronomy and Space Science and Key Laboratory for Modern Astronomy and Astrophysics, Nanjing University, Nanjing 210023, China} 

\author[0000-0002-7153-4304]{Chun Xia}
\affil{School of Physics and Astronomy, Yunnan University, Kunming 650050, China}

\author[0000-0003-3544-2733]{Rony Keppens}
\affil{Centre for mathematical Plasma Astrophysics, Department of Mathematics, KU Leuven, Celestijnenlaan 200B, B-3001 Leuven, Belgium}

\author{Kai E. Yang}
\affil{Institute for Astronomy, University of Hawai'i at M\=anoa, 34 Ohia Ku Street, Pukalani, HI 96768, USA}

\begin{abstract}

Solar eruptive activities could occur in weak magnetic field environments and over large spatial scales, especially relevant to eruptions involving intermediate or quiescent solar filaments. To handle the large scales, we implement and apply a flux rope embedding method using regularized Biot-Savart laws in the spherical coordinate system. Combined with a potential field source surface model and a magneto-frictional method, a nonlinear force-free field comprising a flux rope embedded in a potential field is constructed. Using the combined nonlinear force-free field as the initial condition, we then perform a zero-$\beta$ data-constrained magnetohydrodynamic (MHD) simulation for an M8.7 flare at 03:38 UT on 2012 January 23. The MHD model reproduces the eruption process, flare ribbon evolution (represented by the quasi-separatrix layer evolution) and kinematics of the flux rope. This approach could potentially model global-scale eruptions from weak field regions. 

\end{abstract}

\section{Introduction} 

Solar eruptive activities in the solar atmosphere, such as coronal mass ejections (CMEs), flares and filament/prominence eruptions, are the sources of disastrous space weather. Observing and modeling these eruptions are an effective means to understand, predict, and reduce their adverse impacts. Through long-lasting studies in the past, a type of magnetic field configuration, either a flux rope or sheared arcade, can become unstable and erupt, driving various aforementioned solar activities \citep{2011Chen,2018Green,2019Toriumi}. A flux rope is a bundle of magnetic field lines winding around an axis. If the twist is low, say, less than 1 turn, or the toroidal flux is much larger than the poloidal flux, the bundle of field lines becomes a sheared arcade \citep{2020Patsourakos}. Local coronal plasma may cool down and condense on a flux rope or sheared arcade, which appears as a filament or prominence \citep{2020Chen}. A successful filament/prominence eruption is usually accompanied by a CME and flare with the erupting magnetic field and plasma, which involves an induced shock, more complex electric currents, and accelerated particles.

Many studies revealed that magnetohydrodynamic (MHD) instabilities and magnetic reconnection play important roles in triggering and driving flux rope eruptions \citep{1999Antiochos,2000Chen,2001Moore,2005Torok,2006Kliem,2014Keppens,2017Ishiguro}. MHD instabilities and reconnection are regarded as ideal and resistive processes, respectively. The origin of the MHD instabilities is frequently the Lorentz force, and that of the reconnection is the electron and ion dynamics and electric current dissipation mechanism. A powerful tool to study the realistic triggering and driving processes is data-driven or data-constrained MHD models \citep{2013Kliem,2016Jiang,2018Amari,2018Inoue,2019Guo1,2021Zhong}. These models incorporate observations to constrain the initial and boundary conditions, and can reproduce many key observational features, such as the morphology of flare ribbons, kinematics of erupting filaments, and so on \citep{2022Jiang}.

Depending on the locations of their source regions, solar filaments are divided into three categories, namely, active region, intermediate, and quiescent ones \citep{1998Engvold}. An intermediate filament has one footpoint rooting in an active region and the other footpoint in a quiet region. Traditional nonlinear force-free field (NLFFF) extrapolation techniques need extra flux rope embedding strategies, as they will not easily produce a proper flux rope model for an intermediate or quiescent filament as that for an active region filament. Since the magnetic field under an intermediate or quiescent filament is usually not very strong and they mostly detach from the photosphere, the transverse components of the vector magnetic field are too weak to constrain highly twisted features in the corona. For example, statistics conducted by \citet{2017Ouyang} found that 96\% of quiescent filaments are supported by twsited flux ropes, while only 4\% of them are supported by sheared arcades. However, as for active-region filaments, 40\% of them are supported by sheared arcades. There are only a few models that have reconstructed flux ropes in weak field regions \citep{2012Su,2014Jiang,2020Mackay}. The regularized Biot-Savart laws (RBSL) proposed by \citet{2018Titov} provide a recipe to address the issue. The key idea is to embed a magnetic flux rope ($\mathbf{B}_\mathrm{FR}$) into a potential field ($\mathbf{B}_\mathrm{P}$). In this case, the combined magnetic field is not force-free, since the Lorentz force component exerted by $\mathbf{B}_\mathrm{P}$ on the flux rope generally cannot balance the hoop force of the flux rope due to its curvature. It should be further relaxed to a force-free state by, e.g., a magneto-frictional method \citep{2016Guo2,2016Guo1} or a zero-$\beta$ MHD relaxation method \citep{2021Titov,2022Kucera}.

\citet{2019Guo2} implemented the RBSL method in the Message Passing Interface Adaptive Mesh Refinement Versatile Advection Code \citep[MPI-AMRVAC;][]{2003Keppens,2012Keppens,2021Keppens,2023Keppens,2014Porth,2018Xia}, and thus far used it in a Cartesian coordinate system. For example, \citet{2021Guo} simulated a flux rope eruption using the RBSL model embedded in a potential field as the initial condition. Here, we extend the RBSL model to the spherical coordinate system and present a data-constrained zero-$\beta$ MHD model to study the eruption of an intermediate filament. The motivating observation is described in Section~\ref{sec:observation}. Our MHD simulation setup is described in Section~\ref{sec:mhd}. The results are presented in Section~\ref{sec:result}. A summary and discussion are provided in Section~\ref{sec:summary}.

\section{Observation} \label{sec:observation}

In this study, we focus on an M8.7 flare starting at 03:38 UT on 2012 January 23 in active region 11402, which is located in an active region cluster comprising three other regions (active regions 11401, 11405 and 11407). The coronal magnetic field of this active region cluster was extrapolated before by \citet{2012Guo} with the optimization method \citep{2000Wheatland,2004Wiegelmann}. \citet{2016Guo2} developed another NLFFF model for this active region cluster using the magneto-frictional method \citep{1986Yang} developed in MPI-AMRVAC. A flux rope was found in active region 11402. However, if we compare carefully the constructed flux rope with observations by the Atmospheric Imaging Assembly \citep[AIA;][]{2012Lemen} aboard the Solar Dynamics Observatory \citep[SDO;][]{2012Pesnell}, as shown in Figure 9 of \cite{2016Guo2}, the length of the rope is much shorter than the observed filament. To improve on this aspect, we try to make a new model in this study. Moreover, we want to simulate the dynamic process in active region 11402 that causes the M8.7 flare.

\begin{figure}
    \centering
    \includegraphics[width=0.9\textwidth]{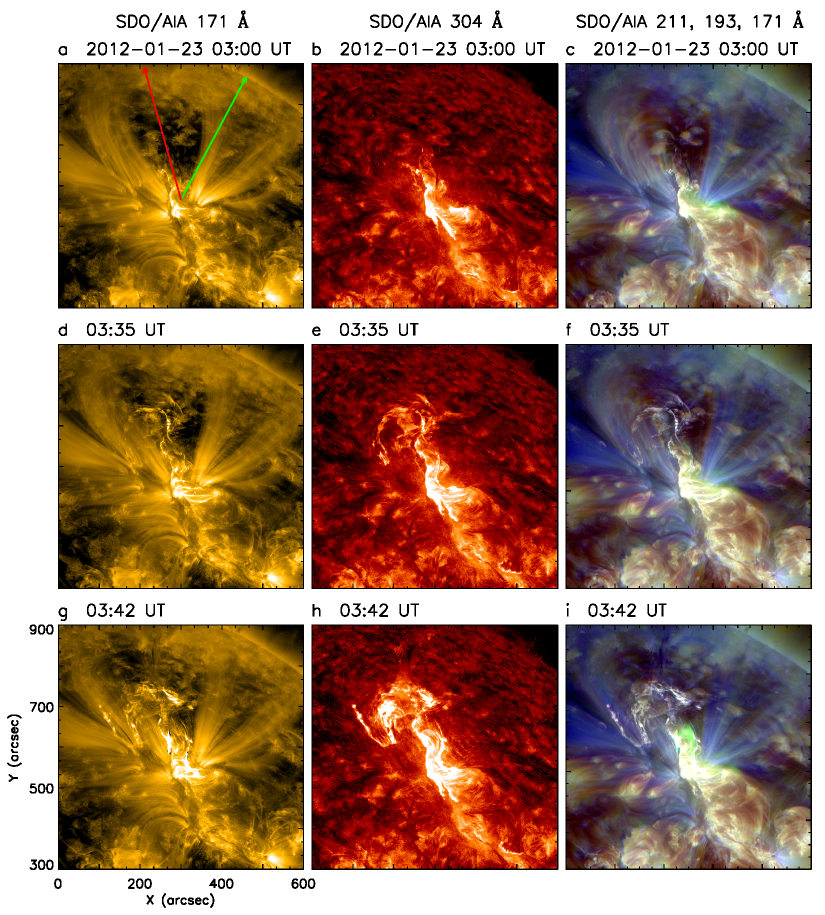}
    \caption{Evolution of the M8.7 flare on 2012 January 23 in different SDO/AIA EUV wavebands. Panels (a), (d) and (g) show the 171 \AA \ images at 03:00, 03:35 and 03:42 UT, respectively. The red arrow in panel (a) points to the  apex of the coronal loops. The green arrow points to the local radial direction projected on the plane of sky. Panels (b), (e) and (h) show the 304 \AA \ images. Panels (c), (f) and (i) show composite images constructed by 211, 193 and 171 \AA \ observations representing the red, green and blue channels, respectively. An animation is attached to this figure, which shows the high cadence evolution of the M8.7 flare in the same wavebands covering a larger time range from 03:00 to 05:00 UT on 2012 January 23.}
    \label{fig:euv}
\end{figure}

Figure~\ref{fig:euv} shows the evolution of the M8.7 flare on 2012 January 23. A prominent filament is suspended in the corona at 03:00 UT (Figures~\ref{fig:euv}a, \ref{fig:euv}b, and \ref{fig:euv}c), well before the flare starts at 03:38 UT. There are some coronal loops overlying the filament as shown in the SDO/AIA 171 \AA \ image (Figures~\ref{fig:euv}a). These loops lean towards the east compared to the radial direction in the active region, which is towards the west as indicated by the red and green arrows in Figure~\ref{fig:euv}a. The filament starts to lift up at 03:35 UT, and brightened flare ribbons can be seen in different EUV wavebands (Figures~\ref{fig:euv}d, \ref{fig:euv}e and \ref{fig:euv}f). At 03:42 UT, the erupting filament has reached and stretched the visible overlying coronal loops (Figures~\ref{fig:euv}g, \ref{fig:euv}h and \ref{fig:euv}i). The eruption process has been analyzed by previous observational studies \citep{2013Cheng,2013Joshi,2014Sterling}. 

\section{MHD Simulation} \label{sec:mhd}

Observations show that the pre-eruptive phenomenon of the source region is dominated by an intermediate filament. We first construct a flux rope model for the filament using the RBSL method. Then, we embed the RBSL flux rope into a potential field derived by the PFSS model. The combined RBSL and PFSS magnetic field is relaxed by the magneto-frictional method to derive an NLFFF model, which further serves as the initial condition for the following zero-$\beta$ MHD simulation.

\subsection{Regularized Biot-Savart Laws in Spherical Coordinates} \label{sec:rbsl}

\citet{2018Titov} proposed the RBSL method for constructing a flux rope $\mathbf{B}_\mathrm{FR}$ with an axis path $\mathcal{C}$, minor radius $a$, magnetic flux $F$, and electric current $I$:
\begin{eqnarray}
\mathbf{B}_\mathrm{FR}=\nabla \times \mathbf{A}_{I} + \nabla \times \mathbf{A}_{F}, \label{eqn:rbsl1} \\
\mathbf{A}_{I}(\mathbf{x})=\frac{\mu I}{4\pi}\int_{\mathcal{C}\cup \mathcal{C}^*}K_{I}(r)\mathbf{R}'(l)\frac{dl}{a(l)}, \label{eqn:rbsl2} \\
\mathbf{A}_{F}(\mathbf{x})=\frac{F}{4\pi}\int_{\mathcal{C}\cup \mathcal{C}^*}K_{F}(r)\mathbf{R}'(l)\times\mathbf{r}\frac{dl}{a(l)^2}, \label{eqn:rbsl3}
\end{eqnarray}
where $\mathbf{x}$ is the position vector of the magnetic field, $l$ is the arc length of the axis path, $\mathbf{R}(l)$ is the position vector of the axis path, $\mathbf{r} \equiv \mathbf{r}(l) = (\mathbf{x} - \mathbf{R}(l))/a(l)$ is the separating vector from the source point $\mathbf{R}(l)$ to the field point $\mathbf{x}$ and normalized by $a(l)$, $\mathbf{R}'=d\mathbf{R}/dl$ is the tangential unit vector, and $\mathcal{C}^*$ is the mirror image of $\mathcal{C}$ referred to the photosphere. For an axial current density distributed parabolically along the flux rope minor radius, the integration kernels are obtained by \citet{2018Titov}:
\begin{eqnarray}
 K_{I}(r)= \begin{cases} \frac{2}{\pi}(\frac{\arcsin r}{r}+\frac{5-2r^{2}}{3} \sqrt{1-r^{2}}) \qquad \text{when} \ 0 \le r \le 1, \\
 \frac{1}{r} \qquad \text{when} \  r > 1, \end{cases} \label{eqn:rbsl4} 
\end{eqnarray}
\begin{eqnarray} 
 K_{F}(r)= \begin{cases} \frac{2}{\pi r^{2}}\left (\frac{\arcsin r}{r}-\sqrt{1-r^{2}}\right )+\frac{2}{\pi}\sqrt{1-r^{2}}
 + \frac{5-2r^{2}}{2\sqrt{6}}\left [1-\frac{2}{\pi} \arcsin\left (\frac{1+2r^{2}}{5-2r^{2}}\right )\right ] \qquad \text{when} \ 0 \le r \le 1,  \\
 \frac{1}{r^{3}} \qquad \text{when} \  r > 1. \end{cases}
 \label{eqn:rbsl4}
\end{eqnarray}
Once the free parameters including the axis path $\mathcal{C}$, minor radius $a$, magnetic flux $F$, and electric current $I$ are determined, we could derive the flux rope magnetic field $\mathbf{B}_\mathrm{FR}$ using Equations~(\ref{eqn:rbsl1})--(\ref{eqn:rbsl3}).

\begin{figure}
    \centering
    \includegraphics[width=0.5\textwidth]{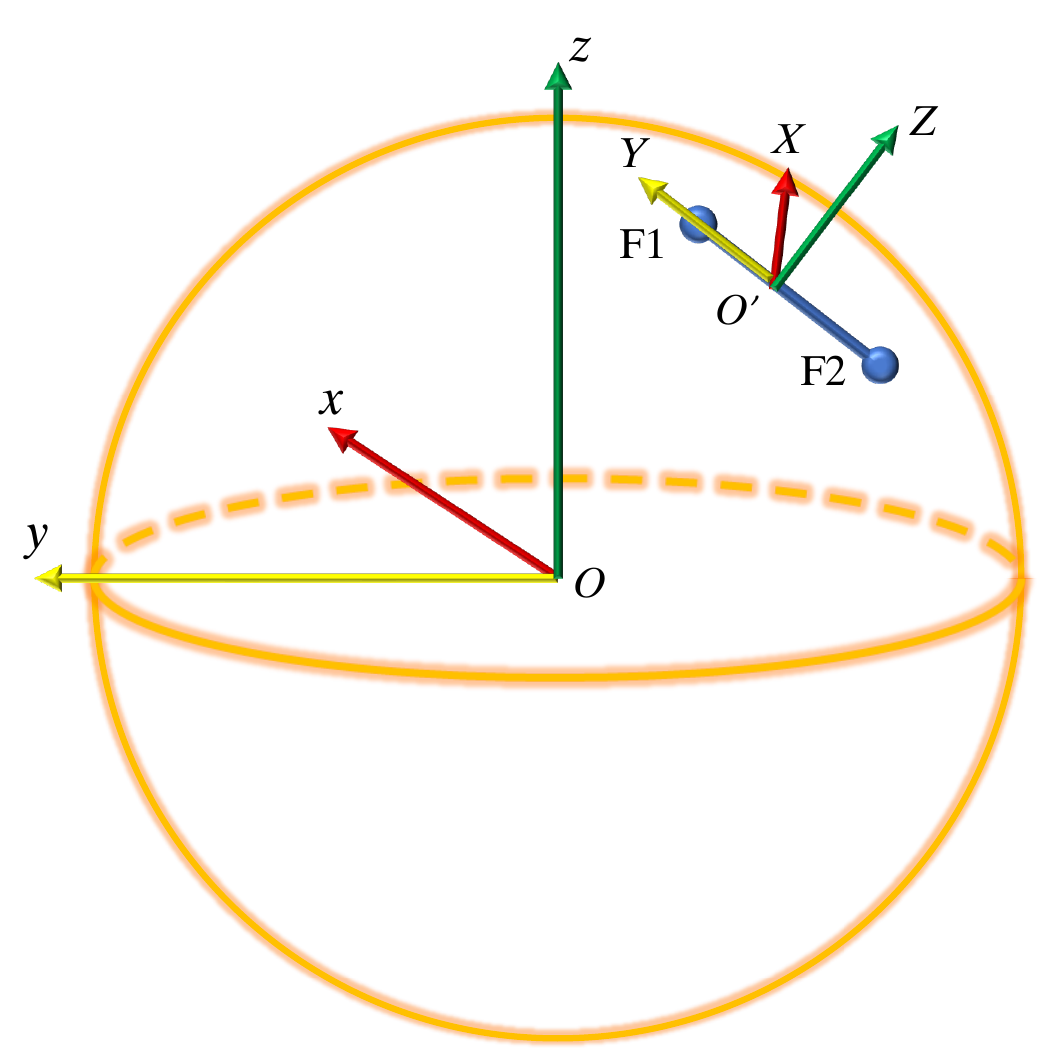}
    \caption{Heliocentric coordinate system and the local coordinate system of the filament. Point $O$ is located at the center of the Sun, the $x$-axis points to the central meridian on the Sun's backside and on the equator, the $y$-axis points to the east solar limb and on the equator, and the $z$-axis constitutes a right-hand system together with the $x$- and $y$-axes. $O'$ is located at the middle point of the two footpoints (indicated by the two blue dots F1 and F2) of the filament on the photosphere, the $X$-axis is perpendicular to both $OO'$ and the line joining the two footpoints, the $Y$-axis is parallel to the line joining the two footpoints, and the $Z$-axis constitutes a right-hand system together with the $X$- and $Y$-axes and pointing radially outward.}
    \label{fig:coord}
\end{figure}

The RBSL method was proposed in the Cartesian coordinate system \citep{2018Titov}. To implement it in a spherical coordinate system, we require that the distance between the two footpoints of the modeled flux rope is as small as possible, though a distance less than or close to the solar radius is tolerable. Therefore, the two footpoints are close enough to keep the radial magnetic field not changing too much on the photosphere when we make the coordinate transformation. We define two coordinate systems, one is the heliocentric $O$-$xyz$ and the other is the local $O'$-$XYZ$ as shown in Figure~\ref{fig:coord}. Since the two coordinate systems are determined, we can convert any useful values between them, such as the coordinates themselves and the magnetic field vecctor components. The idea is to provide the intended spherical coordinates $(r, \theta, \phi)$, which are converted to the coordinates in $O$-$xyz$ and further to $O'$-$XYZ$, then to compute the RBSL magnetic field at these locations in $O'$-$XYZ$, and finally to convert the magnetic field back to $O$-$xyz$ and the spherical coordinates. 

\begin{figure}
    \centering
    \includegraphics[width=0.9\textwidth]{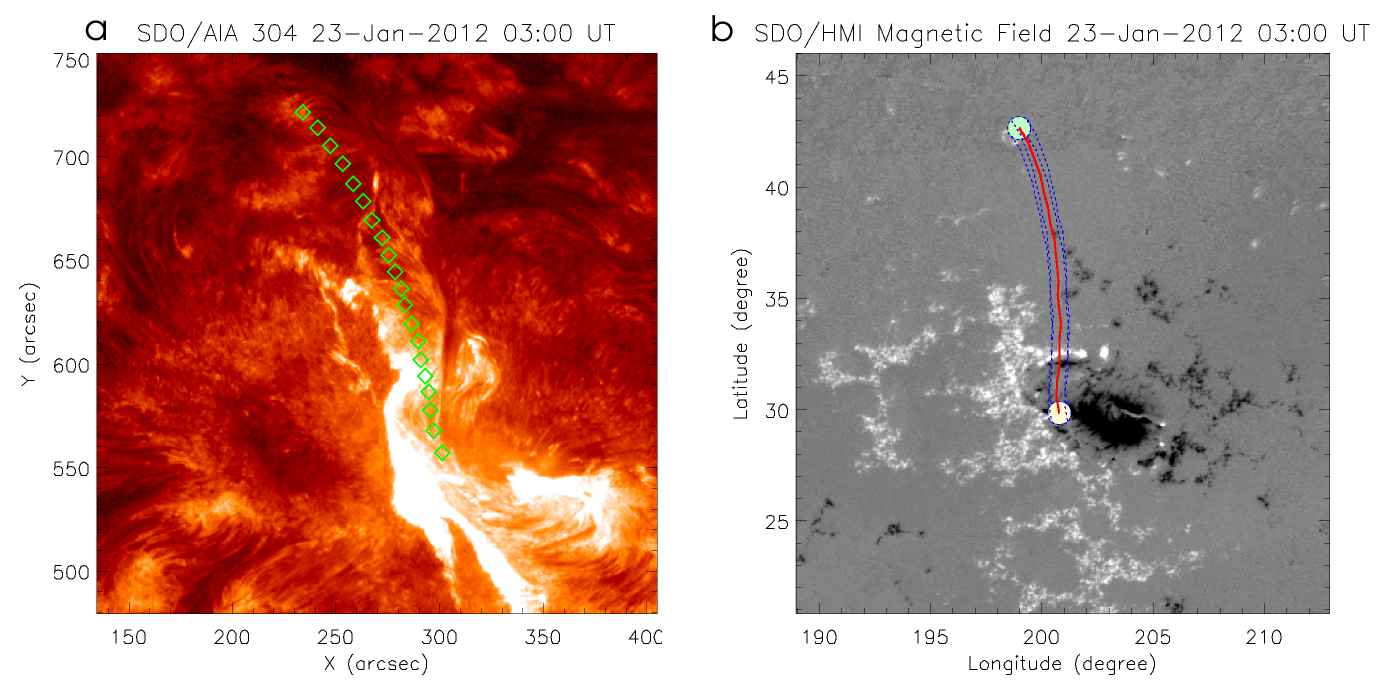}
    \caption{Path of the filament. (a) Filament in the SDO/AIA 304 \AA \ image at 03:00 UT. Green diamonds represent the projected path of the filament body on the solar surface. (b) Projected path overlaid on SDO/HMI magnetic field at 03:00 UT. Green and yellow filled circles represent the two footpoints of the filament. Red and blue lines depict the projected path of the filament on the solar surface.}
    \label{fig:path}
\end{figure}

Computing the RBSL magnetic field in $O'$-$XYZ$ is exactly the same as done in a Cartesian coordinate system \citep{2018Titov}, which has been implemented in MPI-AMRVAC \citep{2019Guo2}. We have to determine the three dimensional (3D) path of the filament $\mathcal{C}$, its minor radius $a$, magnetic flux $F$, and electric current $I$. Figure~\ref{fig:path} shows how the path is determined. Two footpoints are pinpointed first, then the path is depicted along the solar surface by visually following the projection positions of the filament. Note that since the path is a projection of the 3D filament on the solar surface, it does not follow the filament body because of this projection effect (Figure~\ref{fig:path}a). The final 3D path is determined by assigning a height distribution to the projected path, which is a circular arc profile determined by the two footpoints and the apex height, 30~Mm, of the filament. We have to repeat the measurement by trial-and-error, which is 5 times in this case, and compare the constructed 3D flux rope with the observed filament to determine the validity of the selected path. To keep the current circuit closed, we set the mirror image of the observed flux rope path in $O'$-$XYZ$ as the complementary circuit beneath the photosphere. Due to this setting, we use the RBSL kernels derived by \citet{2018Titov} to configure the magnetic flux rope model. Next, the minor radius is estimated by the width of the filament, which is 6 Mm. Then, the axial magnetic flux $F_0$ is estimated by the average of the absolute values of the two footpoints, $|F_+|$ and $|F_-|$, as shown by the green and yellow filled circles in Figure~\ref{fig:path}b. It is found that $F_0 = (|F_+|+|F_-|)/2 = 1.75 \times 10^{20}$~Mx. This flux can be considered as the axial flux of the rope only if the footpoint distance is short. In practice, it is often multiplied by a factor since the method does not require $F=F_0$. In this case, the difference of the longitudinal flux at the footpoints of the flux rope can be compensated by adding the RBSL field and PFSS field together. We finally use $F = 4 F_0$ to simulate the flux rope eruption since it can better match the  evolution of the filament eruption compared to the numerical experiments with other values such as $F = F_0$ and $F = 2F_0$. The electric current is computed by Equation~(12) in \citet{2018Titov}:
\begin{eqnarray}
F=\frac{\pm3}{5\sqrt{2}} \mu Ia, \label{eqn:flux_elec}
\end{eqnarray}
where $\mu$ is the permeability of the plasma. The plus and negative signs are determined by the helicity sign of the active region. If we stand on the positive polarity as shown in Figure~\ref{fig:path}b, the axial field of the flux rope points to the right hand side. Therefore, the chirality of the axial field  is dextral, which implies a negative helicity. Then, we select the negative sign in Equation~(\ref{eqn:flux_elec}).

\subsection{Potential Field Source Surface Model}

The RBSL flux rope constructed with the previous parameters (axis path $\mathcal{C}$, minor radius $a$, magnetic flux $F$, and electric current $I$) is in an internal quasi-equilibrium state. It represents the magnetic field generated by local electric currents in the corona. But it does not include the background coronal magnetic field generated by magnetic polarities on the photosphere other than the flux rope itself. To simulate the magnetic field in the whole active region, we have to embed it in a large-scale magnetic field produced by photospheric magnetogram excluding the flux rope, which could be well approximated by a potential field. It is specifically modelled by the PFSS model in the spherical coordinate system. The PFSS model solves the Laplace equation, which governs the potential field, with the spherical harmonic expansion \citep{1969Altschuler,2003Schrijver}. The solution is represented by the sum of a series of polynomials with harmonic coefficients.

The harmonic coefficients are computed by the radial magnetic field on the photosphere. It is usually derived from a synoptic map constructed with magnetic field observations in a full solar rotation period. Sometimes, a synoptic frame is modified by inserting a snapshot of magnetic field observations into the corresponding synoptic map to increase the temporal resolution on the visible solar disk. In this paper, we adopt the same synoptic frame as used in \citet{2012Guo}, which is constructed by the synoptic map for Carrington rotation 2119 from 2012 January 9 to February 6 and the vector magnetic field observed by the Helioseismic and Magnetic Imager \citep[HMI;][]{2012Scherrer} aboard SDO. However, before using it as the boundary condition for the PFSS model, one more step is needed, i.e., to subtract the radial magnetic field produced by the RBSL flux rope on the two footpoints. According to the design of the RBSL method, the vertical magnetic field $B_Z$ on the local bottom plane $O'$-$XY$ only has values at the two footpoints within minor radius $a$. Note that if the two footpoints are not separated too far, the RBSL radial magnetic field $B_{r,\mathrm{RBSL}}(R_\sun,\theta,\phi)$ at $r = 1 R_\sun$ is also concentrated at the two footpoints. We have to subtract $B_{r,\mathrm{RBSL}}(R_\sun,\theta,\phi)$ from the synoptic frame to prepare the final boundary condition for the PFSS model, which is computed by the PFSS module in MPI-AMRVAC \citep{2014Porth} throughout this paper.

\subsection{Nonlinear Force-Free Field Model}

With the previous two steps, a RBSL flux rope and a PFSS model are derived. We add them together, or equivalently, embed the RBSL flux rope into the PFSS model. This combined magnetic field recovers the radial magnetic field on the photosphere at $r = 1 R_\sun$, but it is not in an exact numerical force-balanced equilibrium state. Indeed, we find that the force-free and divergence free metrics \citep{2000Wheatland} are $\sigma_J = 0.38$ and $\langle |f_i| \rangle = 2.4\times10^{-5}$, respectively. Using the magneto-frictional method built in MPI-AMRVAC \citep{2016Guo2,2016Guo1}, we relax the magnetic field. The bottom boundary is the preprocessed vector magnetic field $B_r$, $B_\theta$, and $B_\phi$ on the photosphere, which are prescribed in the inner ghost layers as the boundary conditions. After iteration of $1.8\times10^5$ steps, the force-free and divergence-free metrics decrease to $\sigma_J = 0.32$ and $\langle |f_i| \rangle = 1.2\times10^{-5}$. We note that $\sigma_J = 0.51$ and $\langle |f_i| \rangle = 9.9\times 10^{-4}$ in \citet{2016Guo2}. Both metrics are improved in this case. This relaxed NLFFF model serves as the initial magnetic field for the following zero-$\beta$ MHD simulations. 

\subsection{Zero-$\beta$ MHD Model}

The zero-$\beta$ MHD model considers three equations, namely, the mass conservation, momentum conservation, and magnetic induction equations \citep{2019Guo1,2021Guo}:
\begin{eqnarray}
\dfrac{\partial \rho}{\partial t} + \nabla \cdot (\rho \mathbf{v}) = 0, \label{eqn:mas} \\
\dfrac{\partial (\rho \mathbf{v})}{\partial t} + \nabla \cdot (\rho \mathbf{v} \mathbf{v} - \mathbf{B}\mathbf{B}) + \nabla (\dfrac{\mathbf{B}^2}{2}) = 0, \label{eqn:mom} \\
\dfrac{\partial \mathbf{B}}{\partial t} + \nabla \cdot (\mathbf{v} \mathbf{B} - \mathbf{B} \mathbf{v}) = -\nabla \times (\eta \mathbf{J}) , \label{eqn:ind}
\end{eqnarray}
where $\rho$, $\mathbf{v}$, and $\mathbf{B}$ are the density, velocity, and magnetic field, respectively, $\eta$ is the resistivity and the electric current $\mathbf{J} = \nabla \times \mathbf{B}$ in the dimensionless form. In the momentum conservation equation, only the Lorentz force is included to change the momentum, while no gas pressure gradient and other external forces (e.g., gravity) are considered. The energy equation is neglected. This model could approximate the evolution of magnetic field in low-$\beta$ circumstances, such as those in the corona, although it neglects some important MHD processes. For example, no slow-mode MHD waves are incorporated, and dynamical aspects like Petschek-type reconnection involving slow-mode shocks can never be resolved. The computation domain is selected as $[r_\mathrm{min}, r_\mathrm{max}] \times [\theta_\mathrm{min}, \theta_\mathrm{max}] \times [\phi_\mathrm{min}, \phi_\mathrm{max}] = [1.00R_\sun, 1.36R_\sun] \times [43.98^\circ, 69.18^\circ] \times [188.94^\circ, 212.94^\circ]$. It is noted that $\theta$ is measured from the north pole and is the complementary angle of the latitude, and $\phi$ is the longitude measured from the central meridian on the backside of the Sun. The computation domain is resolved by a uniform mesh of $360 \times 420 \times 400$ cells.

The initial condition for the magnetic field is the NLFFF model relaxed from the RBSL and PFSS combined magnetic field at 03:00 UT. The initial condition for the density is provided by a stratified atmosphere model, which is very similar to that in \citet{2019Guo1}. Namely, we first prescribe a step-like function for the temperature $T(r)$:
\begin{eqnarray}
T(r)=
  \begin{cases}
    T_0                 & 1.000 R_\sun \leq r < r_0  \\
    k_T (r-r_0) + T_0    & r_0 \leq r < r_1  \\
    T_1                 & r_1  \leq r < 1.360 R_\sun
  \end{cases}, \label{eqn:temp}
\end{eqnarray}
where $T_0=8.5\times10^3$~K, $T_1=1.0\times10^6$~K, $r_0 = 1.005R_\sun$, $r_1 = 1.014R_\sun$, $k_T=(T_1 - T_0)/(r_1-r_0)$. Then, we compute the density distribution $\rho_\mathrm{strat}(r)$ by:
\begin{eqnarray}
\frac{dp}{dr} = -\rho g, \label{eqn:density}
\end{eqnarray}
with $p=\rho T$ in the dimensionless form. The density at $1 R_\sun$ is assumed to be $\rho(R_\sun) = 1.0\times10^8\rho_0$, where $\rho_0=2.34\times10^{-15}$~g~cm$^{-3}$. Similar to the simulation in \citet{2021Guo}, in order to mimic the slow evolution stage due to the heavy mass of the filament, the initial density at 03:00 UT is increased to $\rho_\mathrm{fil} = 1.0\times10^5\rho_0 = 2.34\times10^{-10}$~g~cm$^{-3}$ wherever the density is smaller than this value, namely, $\rho(r) = \max(\rho_\mathrm{strat}(r), \rho_\mathrm{fil})$. The density $\rho_\mathrm{fil}$ is used to mimic the inertia of the filament material. Although the mass density of filaments still has large range of values due to the uncertainties in measuring hydrogen ionization degree \citep{2010Labrosse}, a typical range is $10^{-14}$--$10^{-12}$~g~cm$^{-3}$ \citep{1986Hirayama,1996Heinzel}. The mass density $\rho_\mathrm{fil}$ is about two orders of magnitude larger than the upper limit of the density range. This is necessary in the zero-$\beta$ model because there is no gravity in the momentum equation. In the second stage between 03:34 UT to 03:44 UT, the initial density at 03:34 UT is reset back to the original density profile $\rho_\mathrm{strat}(r)$ with the minimum density at the coronal top down to $3.4\rho_0$, such a manipulation is aimed to simulate the drainage of filament material. Without the density reset, the slow phase and fast eruption phase of the filament eruption cannot be reproduced with the zero-$\beta$ model. Finally, we note that the initial velocity is always zero everywhere in the computation box.

The boundary condition is a data-constrained case as in \citet{2019Guo1}. The density on all six boundaries is provided by the initial values. The velocity on all six boundaries is zero. The magnetic field in the inner ghost layer on the bottom boundary is the observed vector magnetic field $B_r$, $B_\theta$, and $B_\phi$ at 03:00 UT, and it is provided by a zero-gradient extrapolation for all the other boundaries. To control the magnetic reconnection and the evolution process of the flux rope, we adopt an artificial resistivity in the first two minutes of the simulation (03:00--03:02 UT), where $\eta = \eta_0[(J-J_c)/J_c]^2$ wherever $J \geqslant J_c$. We set $\eta_0 = 8.1\times10^{12}$~cm$^2$~s$^{-1}$ and $J_c = 1.1\times10^{-8}$~A~cm$^{-2}$. At other times, there is no explicit resistivity, so only numerical dissipation is present. The resistivity is needed at the beginning to decrease the electric current, so the flux rope does not writhe too much during the eruption. It is turned off at later times to keep the flux rope coherent and not to be dissipated away. Finally, we adopt a three-step time integration, HLL Riemamn solver and Koren limiter in MPI-AMRVAC to solve the zero-$\beta$ MHD equations.

\section{Results} \label{sec:result}

\begin{figure}
    \centering
    \includegraphics[width=0.8\textwidth]{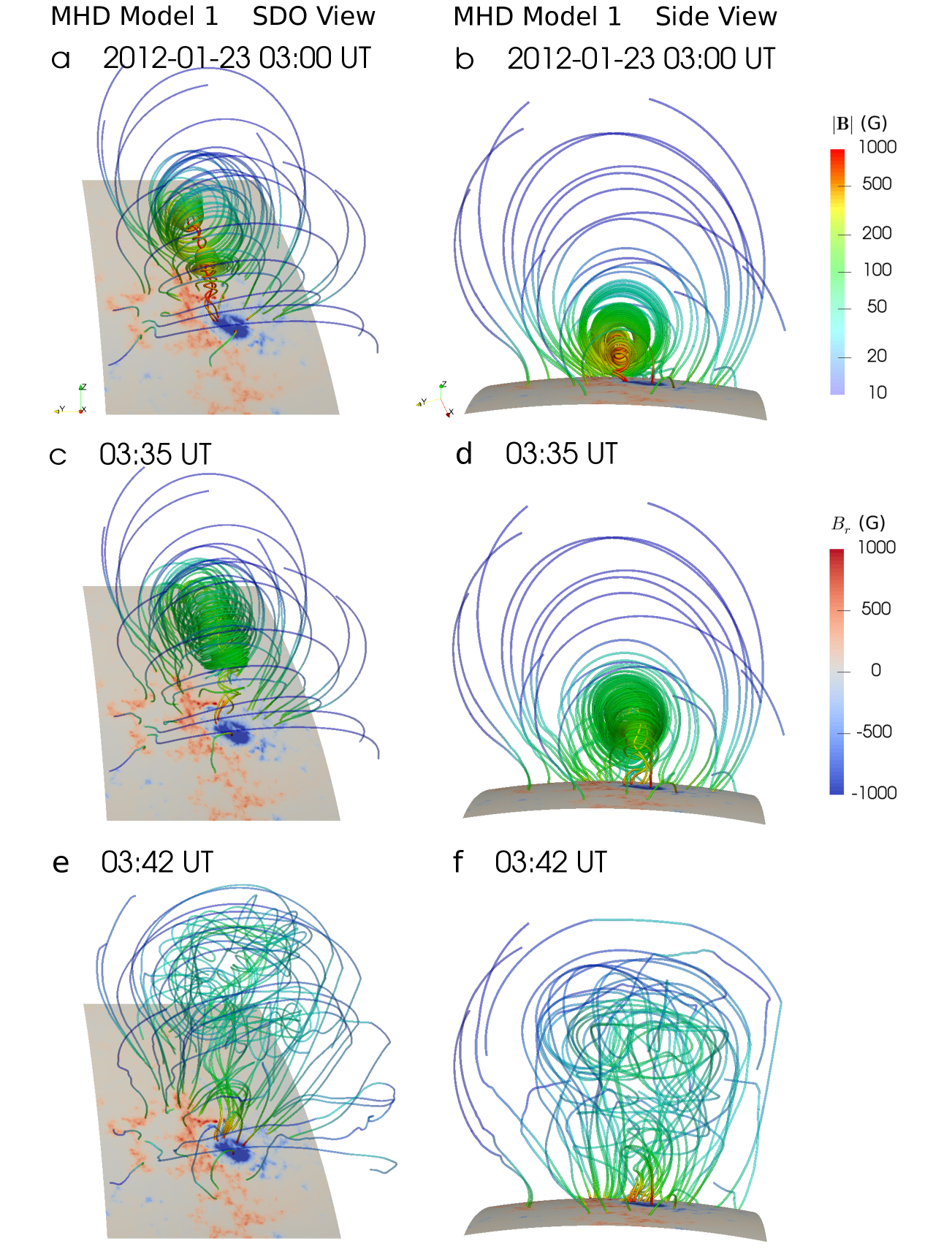}
    \caption{Evolution of magnetic field lines in the MHD simulation. Panels (a), (c) and (e) show the SDO view at 03:00, 03:35 and 03:42 UT, respectively. Panels (b), (d) and (f) show a side view from the south end of the flux rope to the north end. An animation is attached to this figure, showing the evolution of the magnetic field in the time range of 03:00--03:44 UT. The time cadence is 2 minutes between 03:00--03:34 UT, and it is 12 seconds between 03:34--03:44 UT.}
    \label{fig:snapshot}
\end{figure}

Figure~\ref{fig:snapshot} and the attached animation show the evolution of magnetic field lines in the MHD simulation. A magnetic flux rope relaxed from the RBSL and PFSS combined magnetic field suspends in the corona as shown in Figure~\ref{fig:snapshot}a. The south footpoint is rooted in the major negative polarity, and its north footpoint in a quiescent region with a weak positive polarity (see also Figure~\ref{fig:path}b). If we look from a side view as shown in Figure~\ref{fig:snapshot}b, the flux rope displays a cavity morphology as often observed beyond the solar limb. As we reset the density distribution by decreasing the high density to a lower value at 03:34 UT, the flux rope starts to accelerate. The snapshots at 03:35 UT in Figures~\ref{fig:snapshot}c and \ref{fig:snapshot}d are typical ones at the beginning of the acceleration. The body of the flux rope rises and expands radially outward. At 03:42 UT, the flux rope rises up to a certain height, close to the top boundary of the computation box. The flux rope axis writhes a little due to its large twist. 

\begin{figure}
    \centering
    \includegraphics[width=0.8\textwidth]{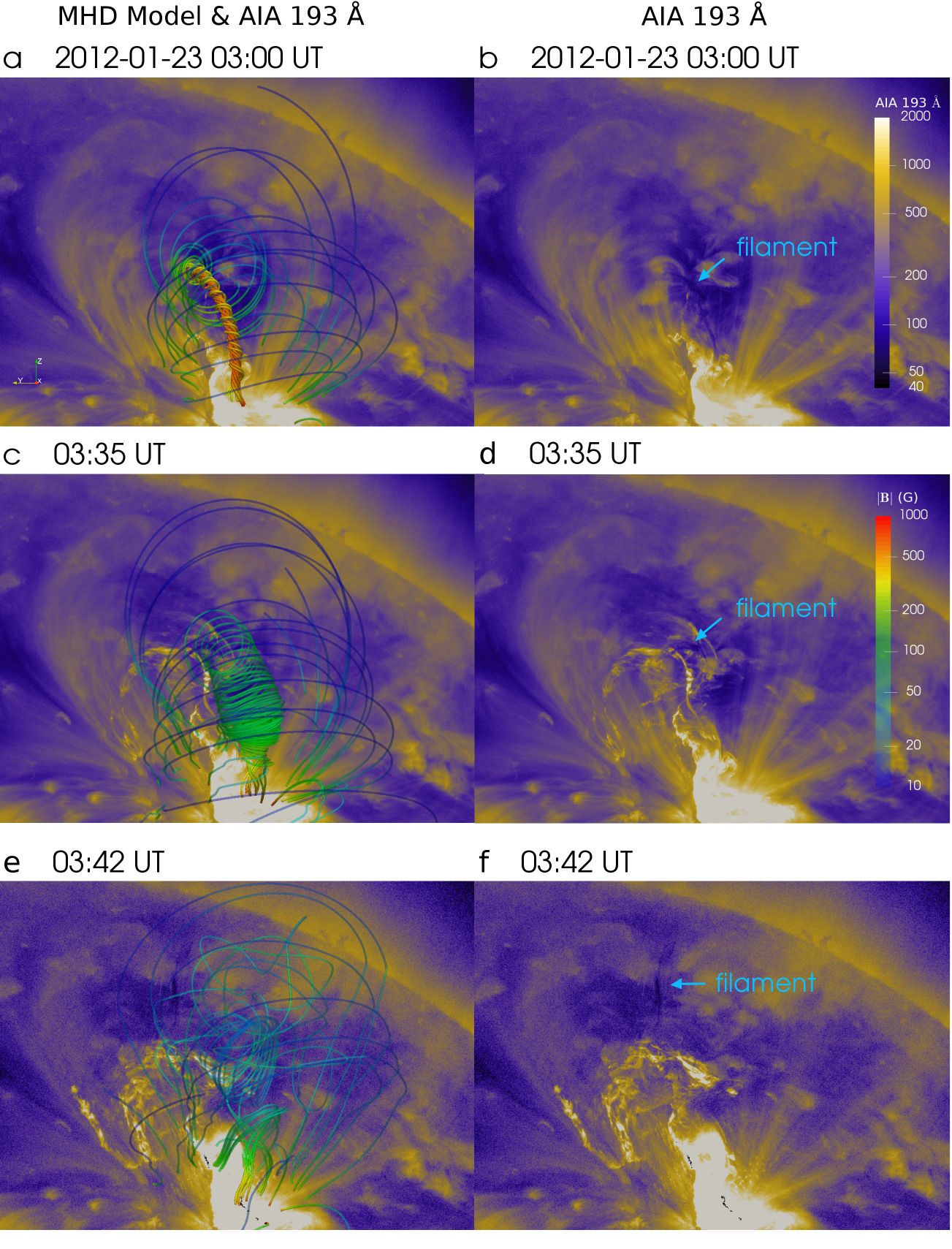}
    \caption{MHD model overlaid on SDO/AIA 193 \AA \ observations. (a, c and e) Background images show the 193 \AA \ observations at 03:00, 03:35 and 03:42 UT from top to bottom panels. Colored lines represent the magnetic field lines in the MHD model. (b, d and f) SDO/AIA 193 \AA \ images highlighting the erupting filament indicated by the cyan arrows at 03:00, 03:35 and 03:42 UT from top to bottom panels. }
    \label{fig:comparison}
\end{figure}

To compare the MHD model with observations, we overlay the magnetic field lines on SDO/AIA 193 \AA \ images as shown in Figure~\ref{fig:comparison}. Since the MHD model is computed in the spherical coordinate system, the alignment between it and SDO/AIA observations is straightforward. We put both the MHD and observational data in the heliocentric coordinate system, $O$-$xyz$. They are aligned in Figure~\ref{fig:comparison}. At 03:00 UT, the magnetic field is our initial condition and the shape of the magnetic flux rope coincides with the filament well, since the path is determined by the filament itself. The good alignment also demonstrates that our method to determine the flux rope path as described in Section~\ref{sec:rbsl} is reliable. At 03:35 UT, the filament starts to erupt, so it shows a bright thread as indicated by the arrow in Figure~\ref{fig:comparison}d. The magnetic flux rope in the MHD model also starts to accelerate (Figure~\ref{fig:comparison}c). Filament matter drains along both the northern and southern legs to their footpoints. At 03:42 UT, the magnetic flux rope has risen outside the field of view (Figure~\ref{fig:comparison}e), and only the filament material in the northern leg could be seen (Figure~\ref{fig:comparison}f).

\begin{figure}
    \centering
    \includegraphics[width=0.8\textwidth]{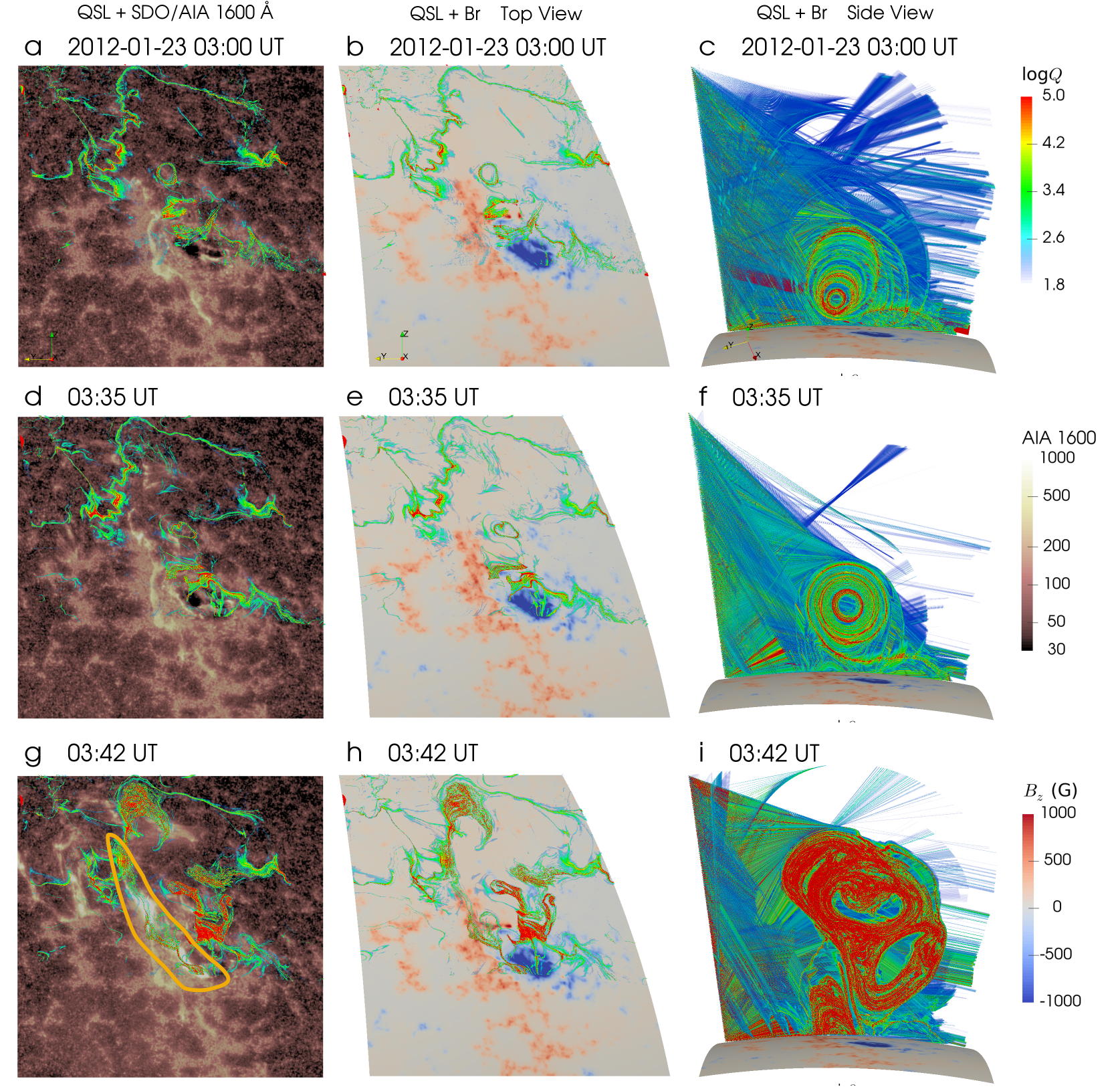}
    \caption{QSL overlaid on SDO/AIA 1600 \AA \ images and SDO/HMI magnetic field. (a, d and g) Distribution of the logarithm of the squashing degree ($\log Q$) on the solar surface overlaid on the SDO/AIA 1600 \AA \ images at 03:00, 03:35 and 03:42 UT from top to bottom panels. The closed orange curve in panel (g) indicates the region where QSLs coincide with flare ribbons. (b, e and h) Distribution of $\log Q$ on the solar surface overlaid on the radial magnetic field observed by SDO/HMI. (c, f and i) Distribution of $\log Q$ on the $\theta=50.28^\circ$ plane. An animation is attached to this figure, showing the evolution of the squashing degree $\log Q$ in the time range of 03:00--03:44 UT. The time cadence is 2 minutes between 03:00--03:34 UT, and it is 12 seconds between 03:34--03:44 UT.}
    \label{fig:qsl}
\end{figure}

Quasi-separatrix layers (QSLs) are 3D thin volumes indicating a drastic change of magnetic field line linkage \citep{1995Priest,1996Demoulin}. QSLs are potential regions for electric current accumulation and magnetic reconnection. \citet{2002Titov} proposed the squashing degree $Q$ to locate QSLs, which are regions where $Q \gg 2$. \citet{2017Tassev} and \citet{2017Scott} proposed new methods using transported displacement vectors or tangents along field lines to calculate the squashing degree $Q$. Kai E. Yang implemented a QSL locator code (K-QSL) using the formulation of \citet{2017Scott}, which is released on GitHub\footnote{https://github.com/Kai-E-Yang/QSL}. One advantage of this code is that it can also be applied to a spherical coordinate system. Therefore, we apply this QSL locator code to our MHD model. Figure~\ref{fig:qsl} and the attached animation show the distribution of the logarithm of the squashing degree ($\log Q$). We find that the erupting flux rope stretches overlying field lines, and new QSLs on the solar surface form during this process as shown in the closed orange curve in Figure~\ref{fig:qsl}g. These new QSLs coincide with the flare ribbons observed by the SDO/AIA 1600 \AA \ waveband. The flux rope is wrapped by QSLs, which is clearly revealed by the $\log Q$ distribution on a $\theta=50.28^\circ$ plane in Figures~\ref{fig:qsl}c, \ref{fig:qsl}f, and \ref{fig:qsl}i. The evolution of $\log Q$ on this vertical plane displays the eruption process and expansion of the flux rope.

\begin{figure}
    \centering
    \includegraphics[width=0.8\textwidth]{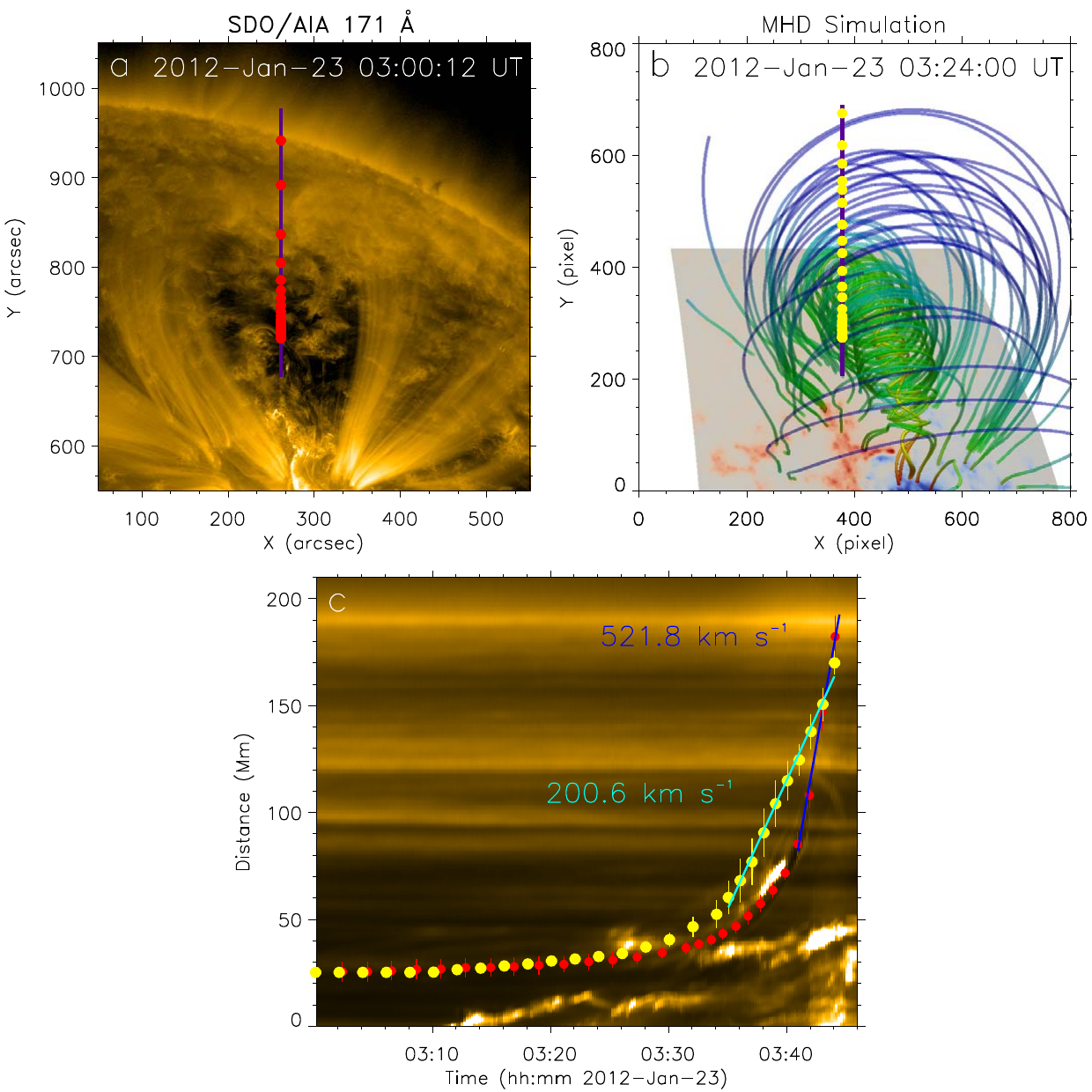}
    \caption{Kinematics of the erupting filament in observation and flux rope in MHD simulation. (a) The purple line over-plotted on the SDO/AIA 171 \AA \ image at 03:00 UT indicates the slice that we choose to measure the time-distance profile of the erupting filament. Red dots are one example of measurements of the filament positions from 03:00--03:44 UT. (b) The purple line over-plotted on the MHD simulation at 03:24 UT indicates the slice that we choose to measure the velocity of the erupting magnetic flux rope. Yellow dots are one example of measurements of the flux rope positions from 03:00--03:44 UT. (c) Red dots and vertical line segments represent the positions and errors measured from the observation of SDO/AIA 171 \AA \ waveband. Yellow dots and vertical line segments represent the positions and errors measured from the MHD simulation. The blue and cyan lines are linear fittings to the time-distance profile of the filament and flux rope at the fast erupting stage, which lead to speeds of 521.8 km~s$^{-1}$ and 200.6 km~s$^{-1}$, respectively. }\label{fig:time_slice}
\end{figure}

We compare the MHD simulation quantitatively with observations by tracking the kinematics of the erupting flux rope and filament. Figure~\ref{fig:time_slice} displays the time-distance profiles for them. We first select a slice vertically in the SDO/AIA 171 \AA \ images and in our MHD simulation (Figures~\ref{fig:time_slice}a and \ref{fig:time_slice}b). The slices are aligned with each other. Then, we measure the positions of the filament and flux rope at different times for both the SDO/AIA 171 \AA \ data series and MHD simulation snapshots. We repeat this measurement ten times for each of them. The average positions and standard deviations are regarded as the final positions and measurement errors as shown in Figure~\ref{fig:time_slice}c. The time-distance profile from the MHD simulation is very similar to that from the observation. Both exhibit two phases in the evolution, including a slow rising phase and a fast erupting phase. The velocity reaches 200.6~km~s$^{-1}$ in the fast erupting phase of the MHD simulation. The evolution in the fast erupting phase of the observation shows a slightly different transition, where the velocity is lower than the simulation at $\sim$03:36 UT but higher than it at $\sim$03:44 UT.

\section{Summary and Discussion} \label{sec:summary}

We implemented and applied a RBSL model in the spherical coordinate system, embeded it in the PFSS model, and relaxed the combined model by the magneto-frictional method in MPI-AMRVAC. The relaxed model resembles the SDO/AIA multi-wavelength observations very well, including the intermediate filament and coronal loops. Using the relaxed model as the initial condition, we performed a zero-$\beta$ MHD simulation for the M8.7 flare on 2012 January 23. The simulation also reproduced the intermediate filament eruption well. The general morphology and eruption direction of the simulation are similar to those in the SDO/AIA observations. The QSLs on the bottom surface display separating flare ribbons, and the QSLs on a vertical surface with $\theta=50.28^\circ$ show the rising and expanding flux rope. The height-time profile measured from the MHD simulation is close to that from the SDO/AIA 171 \AA \ observation, which indicates that the MHD model reproduces the kinematics of the observed filament assuming the magnetic configuration is a flux rope to a good extent.

The average twist of the flux rope is $-3.4$ turns at 03:02 UT, which is computed by the formula in \citet{2006Berger}. We think this magnetic structure is feasible for the studied case, since multi-wavelength emission is usually different from magnetic field structure as noticed by other researchers \citep[e.g.,][]{2017Fan}. High twist flux rope is also prone to appear in intermediate and quiescent prominence regions \citep[e.g.,][]{2015Su,2020Mackay,2021GuoJH}.

There are other MHD simulations in a spherical wedge, where they either use more advanced unstructured adapted mesh \citep{2018Amari}, or sophisticated MHD equations including full thermodynamics \citep{2017Fan}. Our model only uses uniform grid and zero-$\beta$ model. Taking the full advantage of MPI-AMRVAC, we could extend the simulation using adaptive mesh refinement or stretched grids, and including more physics with full thermodynamic MHD equations. Besides, \citet{2018Amari} studied an active region with strong magnetic field, which can be constructed directly by an NLFFF extrapolation. \citet{2017Fan} used a fully parameterized magnetic flux rope model, which is not constrained by observations. The model developed in this work can be applied to intermediate or quiescent filaments with weak magnetic field on the photosphere. Meanwhile, the magnetic field model is constrained by observations. 

When comparing simulations with observations, several features, such as the flux rope and flare ribbons, are related to QSLs, which are quantitatively parameterized by the squashing factor $Q$. Various QSL calculation codes have been developed. Recently, \citet{2022Zhang} compared different QSL codes, all in the Cartesian coordinate system. Locating QSLs in spherical settings has only been done in very few cases, such as \citet{2011Titov}, who used a generalized definition of the squashing factor $Q$ proposed by \citet{2007Titov}. In order to calculate the $Q$ distribution in the spherical coordinates, one of the coauthors of this paper developed an open-source, spherical QSL locator code (K-QSL), based on the method proposed by \citet{2017Scott}. The advantage of this code is that it can be applied to static magnetic field models (e.g., PFSS) and MHD simulations in both the Cartesian and spherical coordinate systems, as demonstrated here.

This paper illustrates that the RBSL model together with the PFSS, magneto-frictional method and MHD models in spherical coordinates provides a solid approach to simulate intermediate and quiescent filaments in a large field of view. With these techniques, we could extend these simulations to larger scales, all the way to the interplanetary space.


\acknowledgments

The observational data are provided by courtesy of NASA/SDO and the AIA and HMI science teams. Y.G., J.H.G, Y.W.N., M.D.D. and P.F.C. are supported by the National Key R\&D Program of China (2022YFF0503004, 2021YFA1600504, and 2020YFC2201200) and NSFC (12127901, 11773016 and 11733003). R.K. received funding from the European Research Council (ERC) under the European Union’s Horizon 2020 research and innovation program (grant agreement No. 833251 PROMINENT ERC-ADG 2018), and from Internal funds KU Leuven, project C14/19/089 TRACESpace and FWO project G0B4521N. The numerical simulations are performed in the High Performance Computing Center (HPCC) at Nanjing University.



\end{document}